\documentclass[11pt]{article}
\usepackage{moriond,graphicx}

\def\be{\begin{equation}}
\def\ee{\end{equation}}
\def\bea{\begin{eqnarray}}
\def\eea{\end{eqnarray}}

\begin{document}
\vspace*{4cm}
\title{NEW QCD RESULTS FROM LEP}

\author{ T. Wengler }

\address{CERN, PH Department, 1211 Geneva 23, Switzerland}

\maketitle\abstracts{ I review recent QCD results from LEP. The
emphasis is on results that represent new studies and on puzzling
disagreements of theory and experiment. Further results are
nevertheless mentioned by reference. The new studies discussed in
more detail are the most precise measurement of unbiased gluon jets
to date, strong evidence of color coherence in 3-jet events, and an,
albeit unsuccessful, search for penta-quarks. As yet unexplained
disagreements are observed in photon-photon collisions for high
momentum charged particle and single jet production, and for the
total cross section of b-quark production.}

The LEP community continues to use the unique data samples collected
at center-of-mass energies on and above the $Z$ peak both to improve
existing results and to carry out new studies. In the former
category a new LEP combined value of the strong coupling constant
has been presented: $\alpha_{s}\left(M_{Z}\right) = 0.1202 \pm
0.0003\left(\mathrm{stat.}\right) \pm 0.0009
\left(\mathrm{exp.}\right) \pm 0.0013 \left(\mathrm{hadr.}\right)
\pm 0.0047\left(\mathrm{theo.}\right)$~\cite{bib:as}. The
fragmentation functions have been measured for quark and gluon
jets~\cite{bib:opal-ff}, and charged particle spectra have been
studied~\cite{bib:fragfunc}. In photon-photon collisions the
exclusive production of particles has been studied for the
interaction of two real photons, and the cross section has been
measured for the interaction of two virtual
photons~\cite{bib:gg-part}.

The emphasis of this review, however, will be on new studies, and on
persisting discrepancies between data and theory in the QCD domain.

\section{New QCD studies from LEP}

\begin{figure}
\begin{center}
\includegraphics[width=0.85\textwidth]{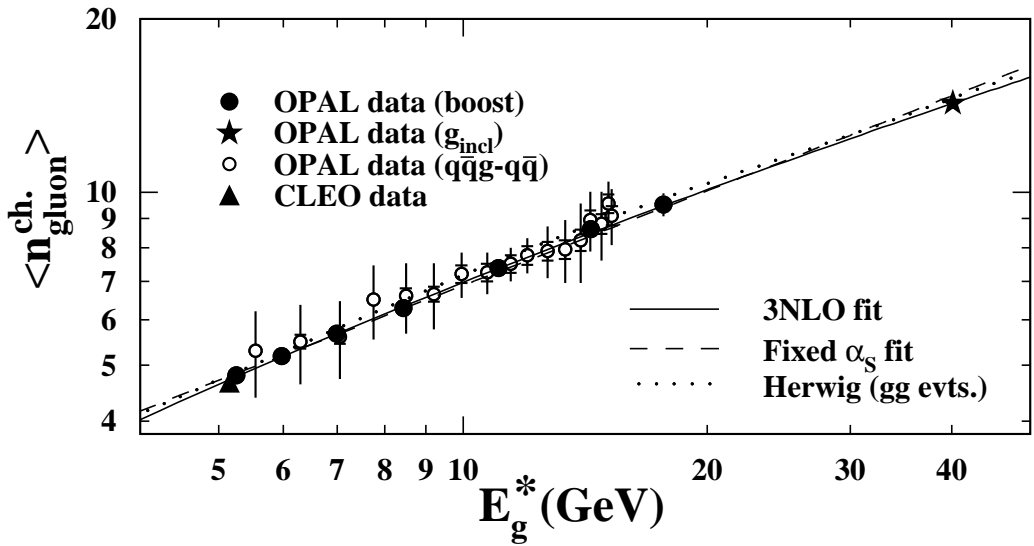}
\caption{The mean charged particle multiplicity value of gluon jets
as a function of the gluon jet energy. The results obtained by OPAL
using the boost algorithm are compared to previous studies and to
theoretical calculations.} \label{fig:unbiased}
\end{center}
\end{figure}

In theoretical studies gluon jets are often defined to be one
hemisphere of a gluon--gluon event. In this case no jet algorithm is
needed to define the jet, which is hence referred to as `unbiased'.
Experimentally jets are mostly defined with the help of jet finding
algorithms. Comparisons to theory then become ambiguous, since the
properties of the gluon jets thus defined may depend strongly on the
algorithm used. These jets are referred to as `biased'. OPAL has
exploited the BOOST algorithm, a new method to construct unbiased
gluon jets, to overcome this problem~\cite{bib:opal-unbiased}. The
method starts from a $Z\rightarrow q\bar{q}g$ event. The partonic
final state is decomposed into a $qg$ and a $\bar{q}g$ dipole, which
are boosted individually to be back-to-back. The dipoles are then
recombined to yield the color structure of a gluon--gluon event.
Experimentally three-jet events are selected using a jet-algorithm
in the sample of hadronic $Z$-decays collected at LEP1. The gluon
jet is identified taking the highest energy jet to be a quark jet,
and by requiring a b-quark tag using standard techniques on the
second or third highest energy jet. The remaining jet is the gluon
jet. The event is now boosted such that the angle $\alpha$ between
the gluon jet and either quark jet is the same. The unbiased gluon
jet is constructed from all particles inside the cone given by
${\alpha}/2$ around the direction of the `biased' jet. Monte Carlo
studies demonstrate that jets thus defined indeed have the same
properties as gluon jets in gluon--gluon events.
Figure~\ref{fig:unbiased} shows the mean charged particle
multiplicity in the unbiased gluon jets as a function of the jet
energy. The data are compared to three previous studies of unbiased
gluon jets, which use rare three-jet events with the two quark jets
almost collinear (OPAL($\mathrm{g_{incl}}$)), a comparison of
two-jet and three-jet events (OPAL($\mathrm{q\bar{q}g-q\bar{q}}$)),
and $\Upsilon \rightarrow \gamma gg$ events in CLEO. The results of
the new study are consistent with previous results, but are the most
precise to date for gluon jet energies between 5.25\,GeV and
20\,GeV. Theoretical results fit the data successfully. Many more
results have been obtained using these unbiased gluon jets,
including the first measurements of the $F_2$ and $F_3$ factorial
moments over an energy range, charged multiplicity ratios of quark
and gluon jets, and gluon jet fragmentation functions. The
fragmentation functions of the unbiased gluon jets have been used
for the first time to extract the strong coupling constant
$\alpha_s$, yielding a value consistent with the world average.

Interference effects are fundamental to quantum mechanical gauge
theories like QCD, but it proves difficult to establish their
existence unambiguously in experimental data. For example,
incoherent fragmentation models are able to describe the data with
similar quality as models including coherence effects, although more
parameters are needed in the former case. DELPHI has carried out a
study of color coherence using two-jet and three-jet events
collected at LEP1~\cite{bib:coherence}. Low energy hadrons produced
inside cones perpendicular to the two-jet axis
($\mathrm{d}\sigma_2$) or three-jet plane ($\mathrm{d}\sigma_3$)
cannot be assigned to a specific jet and have to be treated as
coherent emissions. The two quantities can be related in leading
order QCD by the expression: $\mathrm{d}\sigma_3 =
C_\mathrm{A}/\left(4C_\mathrm{F}\right) \left[ \mathrm{\widehat{qg}}
+ \mathrm{\widehat{\bar{q}g}} - \left(1/N^2_\mathrm{c}\right)
\mathrm{\widehat{q\bar{q}}} \right] \mathrm{d}\sigma_2$, where the
topology dependence is described by the terms inside the square
brackets, which use the antennae functions $\widehat{ij}= 2\,
\mathrm{sin}^2 \left(\theta_{ij}/2\right)$. $\theta_{ij}$ is the
opening angle between two jets. The negative term expresses the
interference effects. A comparison of the data to theoretical
predictions with and without this term shows a strong preference for
interference. To quantify this DELPHI fitted the interference term
scaled by a factor $k$ to $\mathrm{d}\sigma_3/\mathrm{d}\sigma_2$,
assuming $C_\mathrm{A}/C_\mathrm{F} = 9/4$. The result is $k= 1.37
\pm 0.05(\mathrm{stat.}) \pm 0.33(\mathrm{sys.})$ with
$\chi^2/dof=1.2$.

\begin{figure}
\begin{center}
\includegraphics[width=0.45\textwidth]{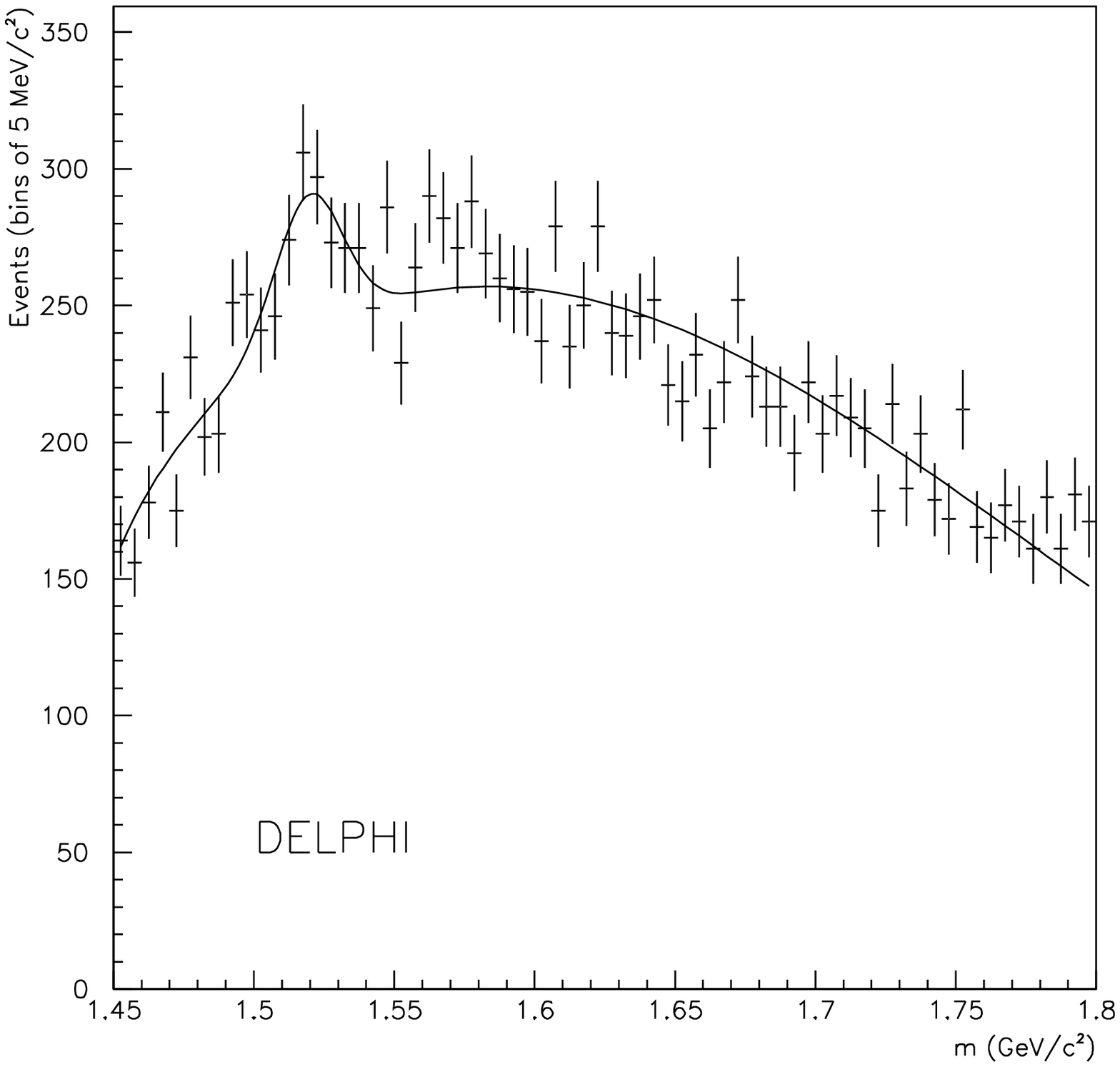}
\includegraphics[width=0.45\textwidth]{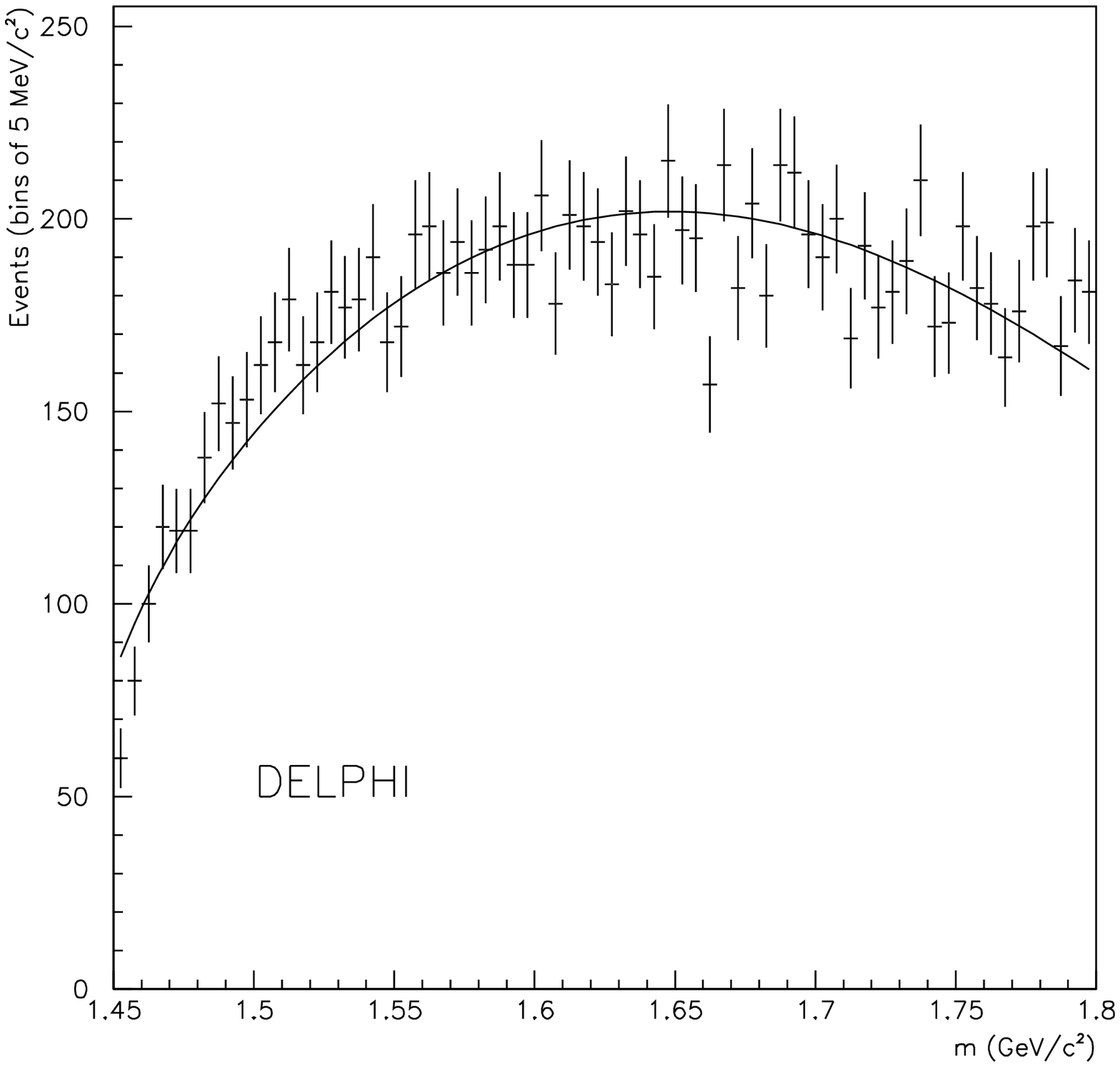}
\caption{The $pK^-$ invariant mass spectrum (left) and the $pK^0$
invariant mass spectrum (right) as observed by DELPHI.}
\label{fig:penta}
\end{center}
\end{figure}

DELPHI has performed a search for a narrow baryonic resonance in the
proton--kaon system, motivated by recent reports of such resonances
with strangeness $S=+1$ with a mass of about 1540 MeV/c$^2$ by
several experiements~\cite{bib:pentaq}. Such a resonance would be a
candidate for a penta-quark state $\Theta^+$. If such a state is
produced at LEP1 like an ordinary baryon, its production rate per
hadronic event should be similar to that of the
$\Lambda\left(1520\right)$, which is $0.0224 \pm 0.0027$. The left
plot of Figure~\ref{fig:penta} shows the $pK^-$ invariant mass
observed by DELPHI in the data taken at the $Z$-peak. A resonance
consistent with the $\Lambda\left(1520\right)$ is observed,
demonstrating the ability of the experiment to reconstruct such
states with the available data sets. No resonance structure is
observed in the $pK^0$ invariant mass spectrum shown on the right
hand side of Figure~\ref{fig:penta}. To extract a limit on the
average multiplicity of the $\Theta^+$ a mass scan is performed in
the region between 1520 MeV/c$^2$ and 1560 MeV/c$^2$. The upper
limit at 95\% C.L. on the average multiplicity of the $\Theta^+$ is
$ \langle N_{\Theta^+}\rangle <0.015$. The $pK^+$ spectrum was
studied in a similar way. Again no resonance structure was observed.
A mass scan in the region between 1500 MeV/c$^2$ and 1750 MeV/c$^2$
yields an upper limit at 95\% C.L. on the average multiplicity of
the $\Theta^{++}$ of $\langle N_{\Theta^{++}}\rangle <0.06$.

\section{Puzzling disagreements of theory and experiments}

\begin{figure}
\begin{center}
\includegraphics[width=0.48\textwidth]{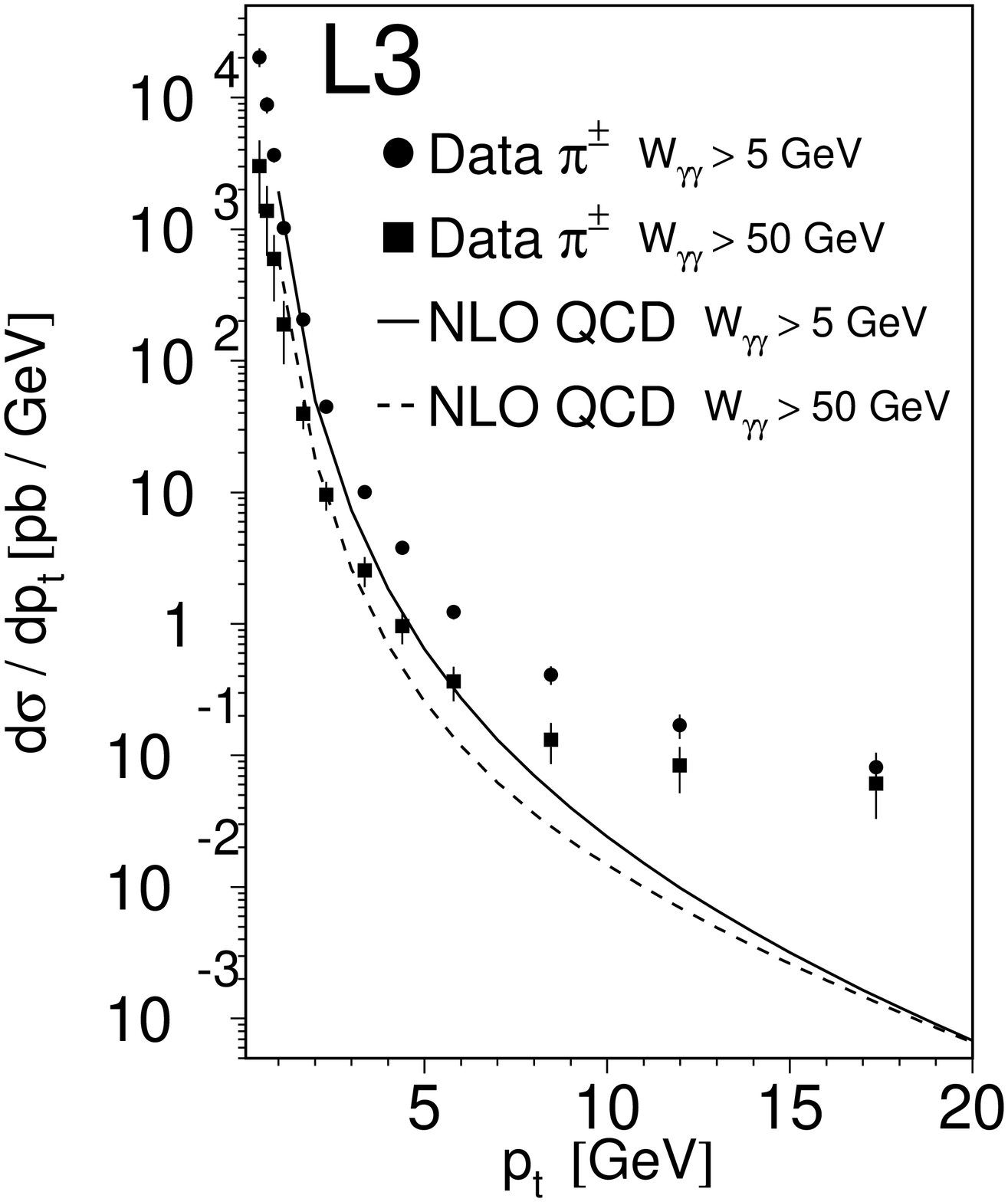}
\includegraphics[width=0.48\textwidth]{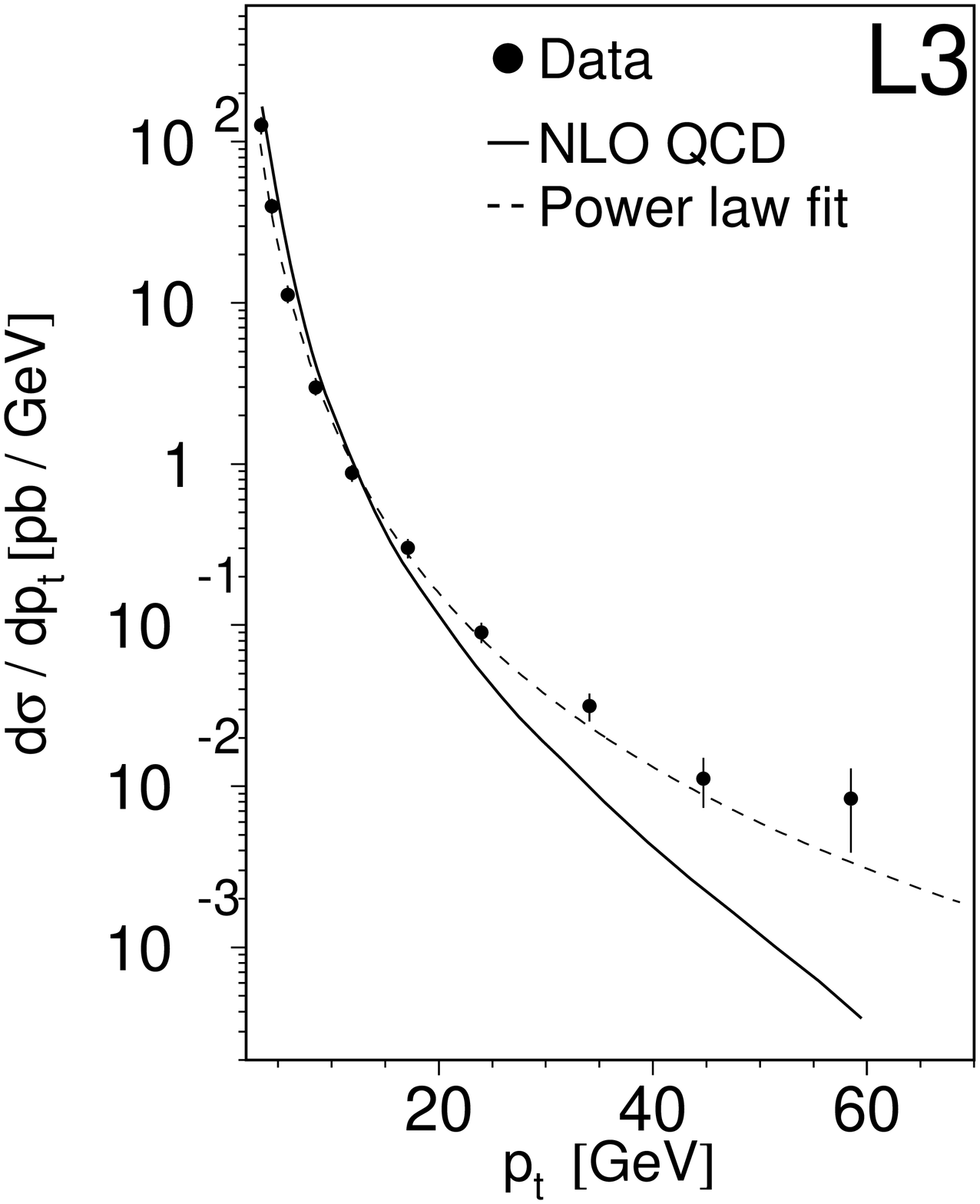}
\caption{The transverse momentum spectra of charged pions for two
values of the invariant mass $W$ of the hadronic system (left) and
the transverse momentum spectrum of single jets (right) as observed
by L3.} \label{fig:jetsparticles}
\end{center}
\end{figure}

QCD has been very successful in recent years in describing high
energy hadronic processes such as jet production in $e^+e^-$, $ep$,
and $p\bar{p}$ reactions. It is all the more important to pinpoint
and study areas where discrepancies persist. Such areas exist for
example in the hadronic interactions of two photons, as studied at
LEP2. L3 has studied the production of charged and neutral pions up
to the highest $e^+e^-$ center-of-mass energies available at
LEP2~\cite{bib:l3particles}. The left plot in
Figure~\ref{fig:jetsparticles} shows the transverse momentum
spectrum of charged pions for two values of $W$, the hadronic
invariant mass of the photon-photon system. In each case it is
evident that the corresponding calculation in next-to-leading order
(NLO) QCD fails to describe the data for momenta larger than about
4\,GeV. At the highest charged particle momenta measured the theory
underestimates the data by more than an order of magnitude. A
similar measurement by L3 of neutral pion
production~\cite{bib:pizero} leads to the same conclusions. Yet the
presence of high momentum particles should indicate the presence of
a hard scale, and the perturbative calculation should be reliable.
The discrepancy can therefore not easily be understood in terms of
the NLO calculation on parton level. Furthermore at high momenta the
interaction of two photons is expected to be dominated by the
so-called direct process, i.e. the exchange of a fermion, such that
uncertainties in the knowledge of the photon structure are not
expected to be very important. To compare the parton level
calculation to the data, it is folded with the appropriate
fragmentation function. While it is not expected that fragmentation
effects could explain discrepancies of this magnitude, it is
interesting to study the same process by different means. L3 has
also measured the production of single jets in photon-photon
collisions~\cite{bib:l3jets}. The right plot of
Figure~\ref{fig:jetsparticles} shows the transverse momentum
spectrum of the jets, again compared to a calculation at NLO QCD. As
in case of the particle spectra, the theory underestimates the data
significantly at high transverse momenta. One has to conclude that
these discrepancies are very significant, and at present not
understood.

Discrepancies between theory and experiment are observed as well for
the total hadronic cross section of b-quark production in
photon-photon collisions. In this case the relatively large mass of
the b-quark should provide the required hard scale for perturbative
calculations. Three measurements now exist using LEP2
data~\cite{bib:gg-bb}. The experimental results are consistent with
each other and are between 2.5 and 4 standard deviations above the
calculation in NLO QCD. While much progress has been made in the
description of b cross sections in $p\bar{p}$ collisions at the
Tevatron~\cite{bib:cacciari}, no solution is in sight yet for
photon-photon collisions. It should be noted however, that large
extrapolations are necessary to extract the total cross section from
the measured distributions. It would be desirable to compare theory
and experiment in a restricted phase space directly accessible
experimentally.


\section*{References}
\begin{thebibliography}{99}

\bibitem{bib:as} M.~Ford, Proceedings of the XXXIXth Rencontres de
Moriond (Electroweak interactons and unified Theories), 20 -- 27
March 2004, La Thuile, Italy.

\bibitem{bib:opal-ff} OPAL Collaboration, G. Abbiendi {\em et al.},
CERN-EP-2004-008, Submitted to Eur.~Phys.~J.~C.

\bibitem{bib:fragfunc} ALEPH Collaboration, A. Heister {\em et al.}, CERN-EP-2003-084,
Submitted to Eur.~Phys.~J.~C.

\bibitem{bib:gg-part} M.N.~Kienzle-Focacci, Proceedings of the Rencontres de physique
de La Vallee d'Aoste, 29 February -- 6 March 2004, La Thuile, Italy.

\bibitem{bib:opal-unbiased} OPAL Collaboration, G. Abbiendi {\em et al.},
Phys. Rev. D69 (2004) 032002.

\bibitem{bib:coherence} DELPHI Collaboration, DELPHI 2004-001, CONF 682.

\bibitem{bib:pentaq} DELPHI Collaboration, DELPHI 2004-002, CONF 683.

\bibitem{bib:l3particles}L3 Collaboration, P. Achard {\em et al.}, Phys.~Lett.~B554 (2003)
105.

\bibitem{bib:pizero} L3 Collaboration, P. Achard {\em et al.}, Phys.~Lett.~B524 (2002) 44.

\bibitem{bib:l3jets} L3 Collaboration, P. Achard {\em et al.}, CERN-EP/2003-055,
Submitted to Phys.~Lett.~B.

\bibitem{bib:gg-bb} L3 Coll., M.~Acciari {\em et al.}, Phys.~Lett.~B503 (2001) 10;\\
$\acute{\mathrm{A}}$.~Csilling (OPAL Collaboration), Proc. of
PHOTON 2000, AIP conf. proc., v. 571 (2000) 276; \\
W. Da Silva (DELPHI Collaboration) Nucl.~Phys.~B 126 (2004) 185.

\bibitem{bib:cacciari} M.~Cacciari, these proceedings.

\end{thebibliography}
\end{document}